\documentclass[a4paper,11pt]{article}

\usepackage[utf8]{inputenc}
\usepackage[english]{babel}
\usepackage{a4wide}
\usepackage[allow-number-unit-breaks=true]{siunitx}
\usepackage{graphicx}
\usepackage{amsmath}
\usepackage{amssymb}
\usepackage{pgf}
\usepackage{xspace}
\usepackage{hyperref}
\usepackage[T1]{fontenc}
\usepackage{authblk}

\title{White Paper:  ARIANNA-200 high energy neutrino telescope}
\author[a]{A.~Anker}
\author[b]{P.~Baldi}
\author[a]{S.~W. Barwick\thanks{sbarwick@uci.edu}}
\author[c]{D.~Bergman}
\author[d]{H. Bernhoff}
\author[e,f]{D.~Z. Besson}
\author[g]{N. Bingefors}
\author[g]{O. Botner}
\author[h]{P. Chen}
\author[h]{Y. Chen}
\author[i,j]{D. Garc\'ia-Fern\'andez}
\author[a]{G. Gaswint}
\author[a,g]{C. Glaser}
\author[g]{A. Hallgren}
\author[k]{J.~C. Hanson}
\author[h]{J.~J. Huang}
\author[l]{S.~R. Klein}
\author[m]{S.~A. Kleinfelder}
\author[h]{C.-Y. Kuo}
\author[a,j]{R. Lahmann}
\author[e]{U. Latif}
\author[h]{T. Liu}
\author[l]{Y. Lyu}
\author[b]{S. McAleer}
\author[h]{J. Nam}
\author[e,f]{A. Novikov}
\author[i,j]{A. Nelles}
\author[a]{M.~P. Paul}
\author[a]{C. Persichilli}
\author[i,j]{I. Plaisier}
\author[h]{J.~Y. Shiao}
\author[n]{J. Tatar}
\author[i]{A.~van Vliet}
\author[h]{S.-H. Wang}
\author[h]{Y.-H. Wang}
\author[i, j]{C. Welling}

\affil[a]{Department of Physics and Astronomy, University of California, Irvine, CA 92697, USA}
\affil[b]{School of Information and Computer Sciences, University of California, Irvine, CA 92697, USA}
\affil[c]{Department of Physics and Astronomy, University of Utah, USA}
\affil[d]{Uppsala University Department of Engineering Sciences, Division of Electricity, Uppsala, SE-75237 Sweden}
\affil[e]{Department of Physics and Astronomy, University of Kansas, Lawrence, KS 66045, USA}
\affil[f]{National Research Nuclear University MEPhI (Moscow Engineering Physics Institute), Moscow 115409, Russia}
\affil[g]{Uppsala University Department of Physics and Astronomy, Uppsala, SE-75237, Sweden}
\affil[h]{Department of Physics and Leung Center for Cosmology and Particle Astrophysics, National Taiwan University, Taipei 10617, Taiwan}
\affil[i]{DESY, 15738 Zeuthen, Germany}
\affil[j]{ECAP, Friedrich-Alexander-Universit\"at Erlangen-N\"urnberg, 91058 Erlangen, Germany}
\affil[k]{Whittier College Department of Physics, Whittier, CA 90602, USA}
\affil[l]{Lawrence Berkeley National Laboratory, Berkeley, CA 94720, USA}
\affil[m]{Department of Electrical Engineering and Computer Science, University of California, Irvine, CA 92697, USA}
\affil[n]{Research Cyberinfrastructure Center, University of California, Irvine, CA 92697 USA}

\date{April 2020}

\begin{document}

\maketitle

\cleardoublepage
\begin{abstract}
The proposed ARIANNA-200 neutrino detector, located at sea-level on the Ross Ice Shelf, Antarctica, consists of 200 autonomous and independent detector stations separated by 1 kilometer in a uniform triangular mesh, and serves as a pathfinder mission for the future IceCube-Gen2 project. The primary science mission of ARIANNA-200 is to search for sources of neutrinos with energies greater than \SI{e17}{eV}, complementing the reach of IceCube. An ARIANNA observation of a neutrino source would provide strong insight into the enigmatic sources of cosmic rays. ARIANNA observes the radio emission from high energy neutrino interactions in the Antarctic ice.  Among radio based concepts under current investigation, ARIANNA-200 would uniquely survey the vast majority of the southern sky at any instant in time,  and an important region of the northern sky,  by virtue of its location on the surface of the Ross Ice Shelf in Antarctica. The broad sky coverage is specific to the Moore's Bay site, and makes ARIANNA-200 ideally suited to contribute to the multi-messenger thrust by the US National Science Foundation, Windows on the Universe – Multi-Messenger Astrophysics, providing capabilities to observe explosive sources from unknown directions. 
The ARIANNA architecture is designed to measure the angular direction to within \SI{3}{\degree} for every neutrino candidate, which too plays an important role in the pursuit of multi-messenger observations of astrophysical sources. 
\end{abstract}

\section{Science enabled by ARIANNA-200}

\begin{figure}[t]
    \centering
    \includegraphics[width=0.79\textwidth]{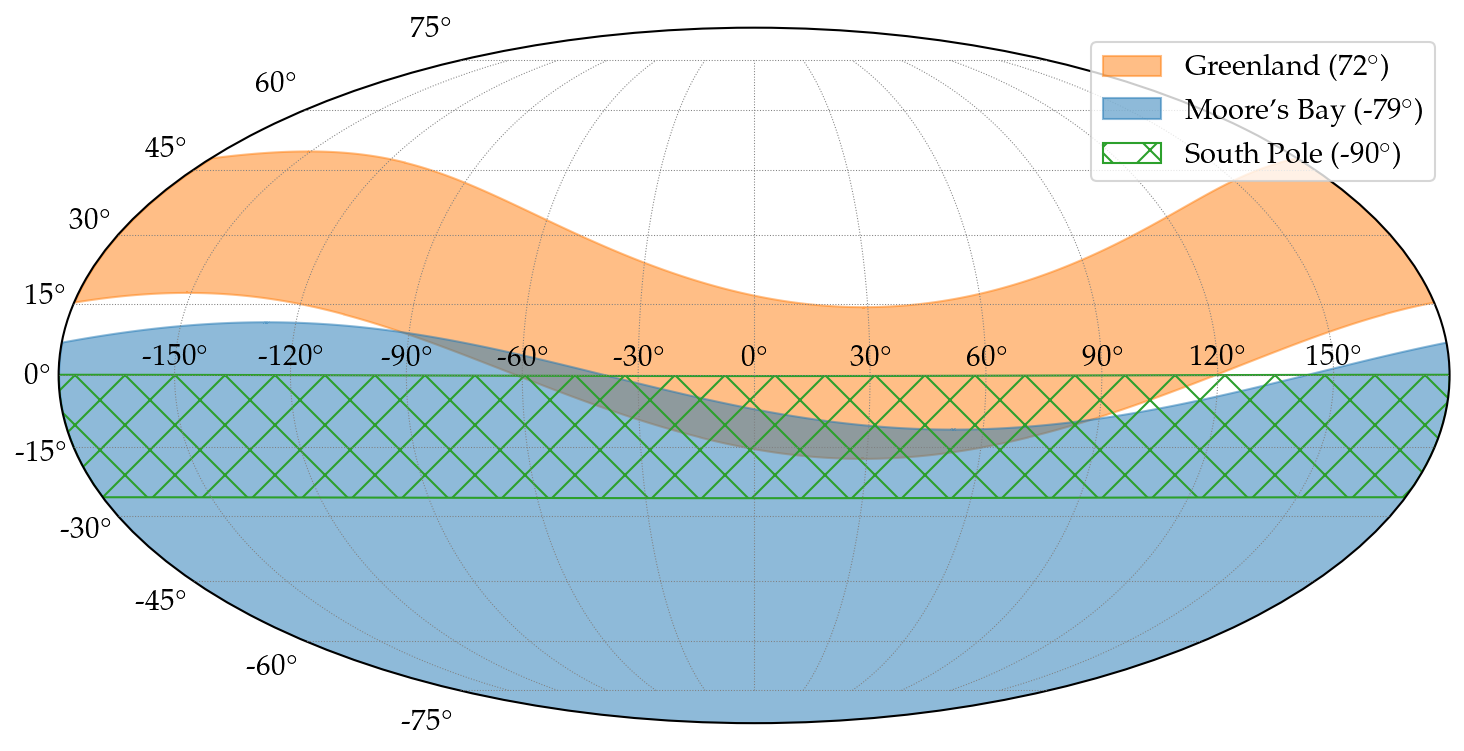}
    \caption{Instantaneous sky coverage of ARIANNA-200 at Moore’s Bay, Antarctica (Blue), plotted in Right Ascension (RA) and Declination (Dec) at one particular time of the day.  For comparison, the sky coverage is shown for radio-based neutrino detectors located at Summit Station in Greenland (gold) and South Pole, Antarctica (green hatch).}
    \label{fig:skycoverage}
\end{figure}

The ARIANNA-200 neutrino detector, located at sea-level on the Ross Ice Shelf, Antarctica, consists of 200 autonomous and independent detector stations separated by 1 kilometer in a uniform triangular mesh. As a consequence of the reflection properties at the ice-water interface at the bottom of the Ross Ice Shelf, ARIANNA-200 views almost the entire southern sky, including the galactic center, with nearly uniform exposure. ARIANNA-200 (Figure \ref{fig:skycoverage}) exceeds the instantaneous sky coverage of all other radio-based neutrino detectors being studied. It's broad sky coverage is ideally suited to contribute to multi-messenger campaigns initiated by gravitational-wave detectors, gamma-ray telescopes, cosmic ray observatories, and neutrino telescopes targeting lower energies such as IceCube \cite{IceCube2013} in the Southern hemisphere, and KM3NeT \cite{Aiello2019} and Baikal-GVD \cite{Baikal2019} in the Northern hemisphere.

\begin{figure}[t]
    \centering
    \includegraphics[width=0.49\textwidth]{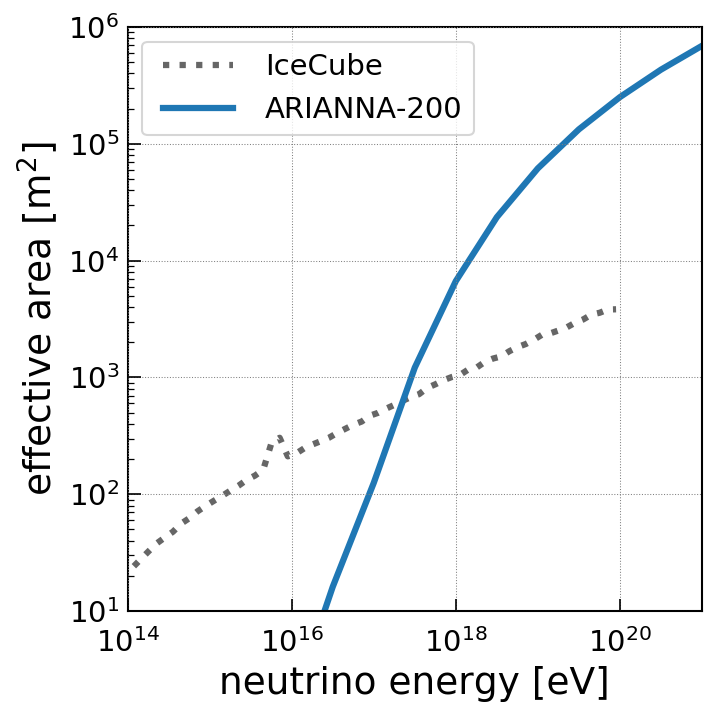}
    \includegraphics[width=0.49\textwidth]{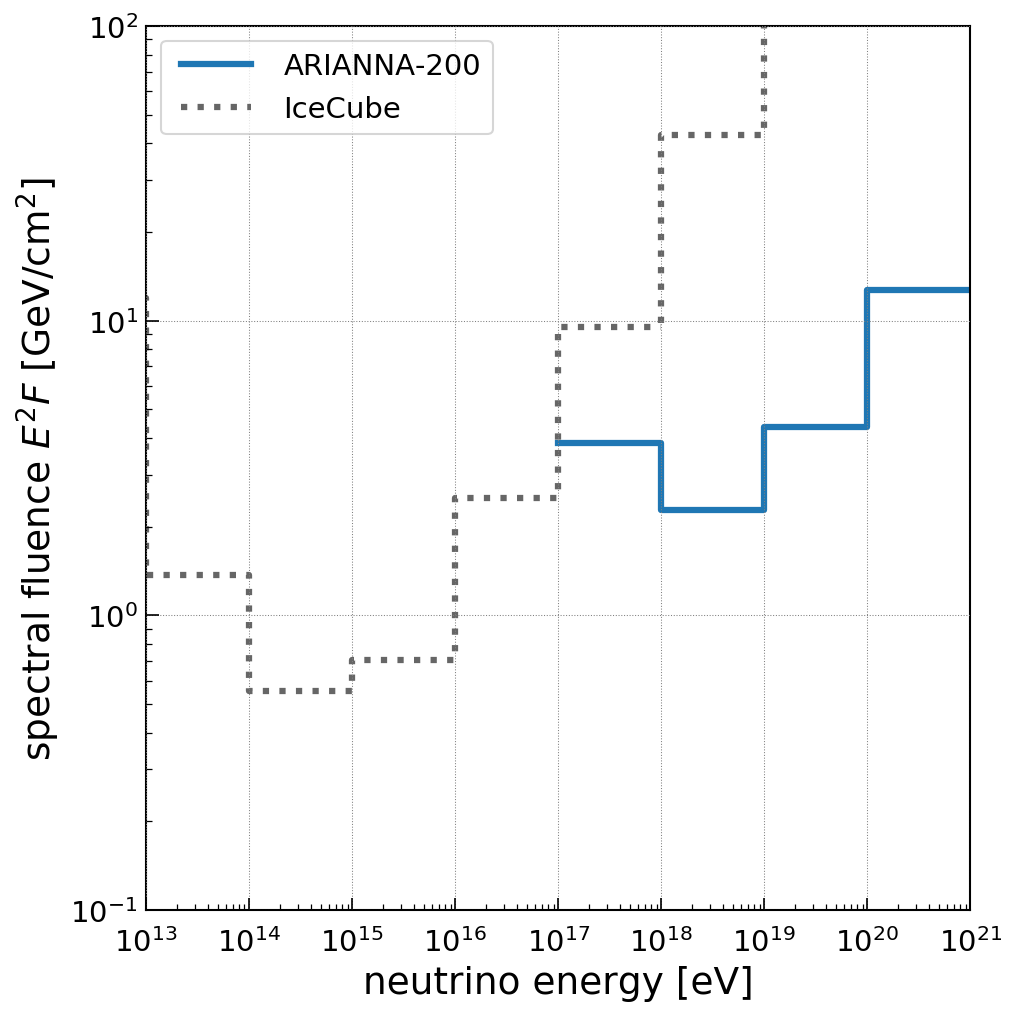}
    \caption{(left) Comparison of effective area for ARIANNA-200 and IceCube as a function of neutrino energy, averaged over neutrino flavor and averaged over the sky.   (right) Comparison of point fluence sensitivity at 90\% C.L. for ARIANNA-200 and IceCube as function of energy, both observing a source at declination of \SI{-23}{\degree}.}
    \label{fig:area}
\end{figure}

The sky coverage of ARIANNA-200 augments the point source capabilities of IceCube. At high neutrino energies ($E_\nu > \SI{\sim e14}{eV}$), the Earth becomes opaque. Thus, at higher energies, both IceCube and ARIANNA-200 observe mostly the Southern sky, leading to a substantial overlap in sky coverage. Figure \ref{fig:area} examines the relative sensitivity as a function of energy for an explosive or flaring source at an arbitrary declination of \SI{-23}{\degree} in the Southern sky. The strong second minimum at \SI{e18}{eV} indicates that ARIANNA-200 will observe about one event for every three sources of the highest energy cosmic rays observed by IceCube, assuming  neutrino production above \SI{e15}{eV} with an unbroken power law up to \SI{e20}{eV} proportional to $E_\nu^{-2}$. A spatially and temporally coincident detection of the same source would establish a hard spectrum up to an energy of \SI{e18}{eV} or greater, and provide a direct link to an accelerator of the very highest energy cosmic rays. A more speculative spectrum proportional to $E_\nu^{-1.8}$ would produce one event in ARIANNA-200 for every neutrino in IceCube with $E_\nu > \SI{e15}{eV}$. 
The model parameter-space for neutrino fluxes of sources is large. Some models suggest that the flux from some neutrino sources may be enhanced at energies close to maximum sensitivity of ARIANNA-200, for example \cite{Fang2019, WaxmanBahcall,FangMetzger2017}, while others predict no observable emission. It is quite possible that new experimental results will be able to guide theory in this respect. 

The simultaneous observation of a point source by IceCube and ARIANNA-200 in different energy ranges would create transformational progress in understanding the half-century old mystery of cosmic rays.  Cosmic rays possess extraordinary high energy, but we do not know the sources of their power, nor the physics responsible for their acceleration. The ARIANNA architecture is designed to measure the angular direction to within \SI{3}{\degree} or better for every neutrino candidate, which too plays an important role in the pursuit of multi-messenger observations of astrophysical sources. Perhaps as few as one neutrino detected by ARIANNA-200, correlated in time and direction with an explosive event observed by IceCube or in some other messenger channel, would provide conclusive steps forward in field of cosmic ray astrophysics.

Apart from the astrophysical neutrinos produced directly at the sources of cosmic rays, cosmogenic neutrinos are produced by the interaction of UHECR protons and cosmic microwave photons \cite{Greisen1966,Zatsepin1966,Berezinky1969,Stecker1979}. These interactions typically still happen close to the source, and the neutrino preserves the cosmic-ray direction. Thus, also cosmogenic neutrinos can reveal the sources of cosmic rays. They have not been detected so far. In 10 years of operation, ARIANNA-200 will be sensitive to cosmogenic fluxes at a level of $E_\nu^2\Phi\le$ \SI{4e-9}{GeV cm^{-2}s^{-1}sr^{-1}}, corresponding to $\sim$10\% of the current limits for neutrino energies above \SI{e18}{eV}, and meets the suggested sensitivity for a IceCube-Gen2 pathfinder mission.  The observation or upper limit from ARIANNA-200 will constrain model parameters, such as source evolution, energy cutoff and cosmic ray composition. 

With a combined fit to the energy spectrum and $X_\mathrm{max}$ distribution (an estimator of the cosmic-ray mass) of UHECR data, the parameters of cosmic-ray sources are estimated from which the cosmogenic neutrino flux can be predicted. However, the analysis is based on a number of simplified assumptions (e.g. a continuous distribution of identical sources and rigidity dependent maximum energies) and the results possess large uncertainties. For example, analysis of the data of the Pierre Auger Observatory located in Argentina \cite{Auger2014, Aab2017} results in substantial differences to an analysis of the data of the Telescope Array (TA) located in Utah \cite{BergmanICRC2019}. The former favors a heavy composition with a low rigidity cutoff at the source resulting in a small cosmogenic neutrino flux, whereas the former favors a high rigidity cutoff and a slightly lighter source composition resulting in a much higher neutrino flux. Furthermore, data of the Pierre Auger Observatory is compatible with an additional proton contribution resulting in substantial increase in the expected neutrino flux \cite{Vliet2019}. 

\begin{figure}[t]
    \centering
    \includegraphics[width=0.6\textwidth]{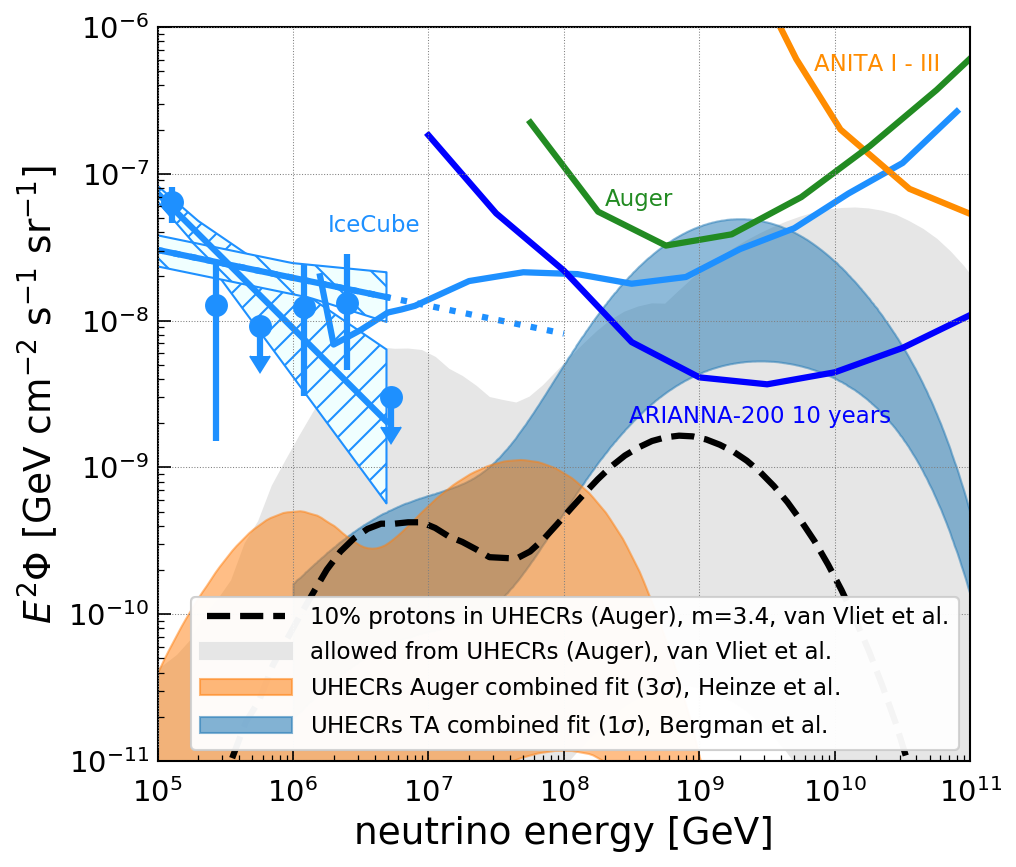}
    \caption{Expected sensitivity of the ARIANNA-200 detector in one-decade energy bins calculated using NuRadioMC \cite{NuRadioMC} for 10 years of operation assuming a uptime of 100\%. Also shown is the measured astrophysical neutrino flux from IceCube using the high-energy starting event (HESE) selection \cite{IceCubeFlux2017} and using a muon neutrino sample \cite{Haack2017}, limits from existing experiments (IceCube \cite{IceCubeFlux2018}, Auger \cite{AugerNeutrinoFlux2015} and Anita \cite{ANITA2018}). The color shaded bands show predictions using a simple astrophysical model with commonly discussed source evolution parameters based on cosmic ray data of the Telescope Array (blue) \cite{BergmanICRC2019, Bergman-Private} and the Pierre Auger Observatory (orange) \cite{Heinze2019}. The dashed line shows a slightly more complex model with an additional small proton component \cite{Vliet2019}. The gray band indicates the range of theoretical uncertainties on model parameters \cite{Vliet2019}. }
    \label{fig:diffuse}
\end{figure}

We summarize the different predictions of cosmogenic neutrinos as well as the predicted ARIANNA-200 sensitivity, and results from existing experiments in Fig.~\ref{fig:diffuse}. The prediction from TA data is well within the reach of ARIANNA-200. For the more pessimistic source parameters derived from Auger data, ARIANNA-200 may observe cosmogenic neutrinos if the proton fraction is larger than 20\% of the total particle number. Thus, ARIANNA-200 will provide new insights into the properties of cosmic-ray sources.

ARIANNA-200 serves as a pathfinder mission for IceCube-Gen2 \cite{IceCubeGen2Whitepaper2019}.  The ARIANNA-200 approach provides a wide $2\pi$ field of view of mostly the Southern Sky, and the largest overlap with IceCube-Gen2 of any location discussed by the community.   The autonomous architecture employed by ARIANNA-200 has successfully operated at the South Pole, and is a viable technological option for devices located at high elevations, such as the South Pole and Greenland.  The surface design of the ARIANNA-200 station provides strong performance in terms of energy and angular resolution, and it provides this level of performance for a large fraction of neutrino candidates \cite{GlaserICRC2019}. 

\section{Detection principle}
Radio emission is generated in ice by particle showers through the Askaryan effect \cite{Askaryan}. If the shower occurs in a dielectric medium, such as ice, the shower develops a time-varying negative charge-excess in the shower front which is primarily due to a collection of electrons from the surrounding medium. The resulting radio emission can be calculated precisely using classical electrodynamics by tracking the movement of the individual particles (see e.g. \cite{ZHSTimeDomain, Alvarez-Muniz2011, ZHAireS2012a, Endpoint2011, AlvarezMuniz2020}) using the well-tested Monte Carlo code ZHS/ZHAireS \cite{ZHS, ZHAireS2012a}. The code incorporates important phenomena such as the LPM effect \cite{Landau:1953um, Migdal:1956tc} that strongly affects the emission for $\nu_e$ charged current interactions. In ice, the electric field increases linearly from MHz frequencies up to a characteristic cutoff of a few GHz. 
Due to coherence effects, the emission is only strong at angles close to the Cherenkov angle, and linearly polarized in the plane defined by the shower direction and propagation direction of the radio signal. Because of this, both the signal arrival direction as well as the polarization need to be measured experimentally to determine the neutrino direction. The observed frequency range is strongest between \SI{100}{MHz} and \SI{1}{GHz} due to properties of the emission and ice attenuation.

The theoretical calculation of the Askaryan emission has been confirmed in accelerator experiments \cite{Saltzberg2001, Gorham2005, Gorham2007, Miocinovic2006, SlacT510}, in particular for showers developing in ice \cite{Gorham2007}. All measurements are consistent with the theoretical prediction within experimental uncertainties. Furthermore, the Askaryan effect has been observed in cosmic ray induced air showers where the Askaryan radiation is subdominant to radio signals emitted by the geomagnetic effect because of the much lower density of air compared to ice \cite{AERAPolarization, LofarPolarization2014}. As it was the case for accelerator measurements, the measurement of the Askaryan radiation from air showers is in agreement with theoretical calculations (e.g. \cite{Scholten2016}).

Radio signals can propagate with little attenuation through ice, allowing an observation of large volumes with a sparse array of detector stations. Field studies at Moore’s Bay measured round trip field attenuation between \SI{300}{m}-\SI{500}{m} in vertical directions \cite{Barrella2011}, and confirmed expectation of excellent reflection from the water-ice interface \cite{Barwick2015}. In-situ radio pulsers at Moore’s Bay \cite{ReedICRC2015} and South Pole \cite{GaswintICRC2019} demonstrated that the direction of the radio emission can be measured to an accuracy of \SI{0.4}{\degree} from typical propagation paths originating from neutrino interactions which probes that the bending of signal trajectories in the firn is well understood. Also the measured polarization is consistent with its expectation, which was obtained from an anechoic chamber measurement of the in-ice pulser, and was reconstructed with a precision of bettern than \SI{3}{\degree}. Also the received signal amplitudes are consistent with expectation after correcting for attenuation from the propagation through the ice \cite{ARIANNAIce2020}. 

In addition, ARIANNA verified its polarization capabilities by measuring cosmic rays which generate radio pulses in the atmosphere with well-characterized polarization. The ARIANNA test bed measured the polarization of these signals to a precision of \SI{7}{\degree} \cite{NellesICRC2019}, providing an independent check on the predicted angular resolution of ARIANNA-200 \cite{GlaserICRC2019}. There is no expectation of significant birefringence, which would complicate a polarization measurement, in the ice at Moore’s Bay, and site studies are consistent with that expectation.

\begin{figure}
    \centering
    \includegraphics[width=0.6\textwidth]{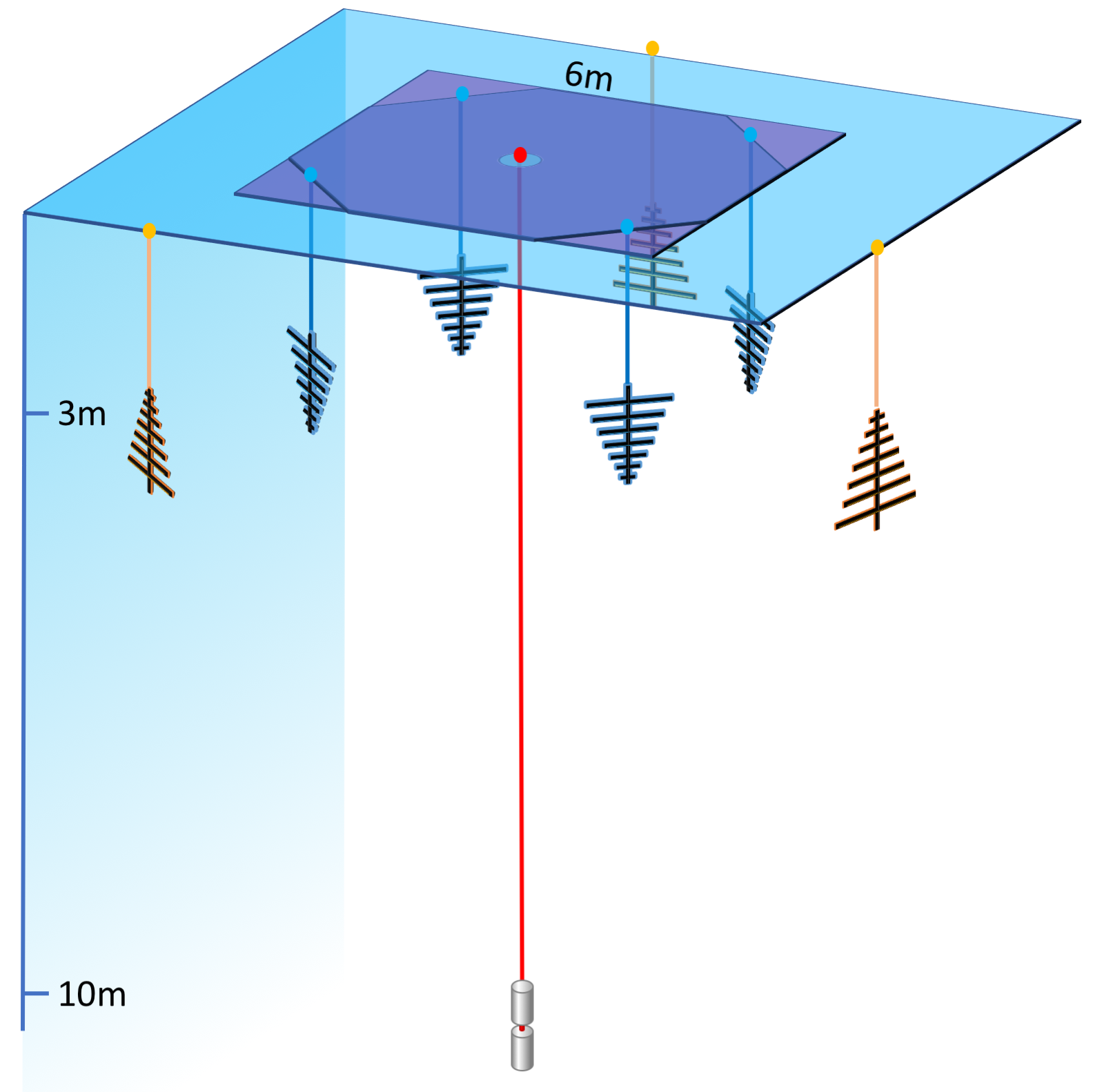}
    \caption{Schematic illustration of one of 200 autonomous, independent stations in the ARIANNA-200 array.  All detector components are within \SI{10}{m} of the surface.  One dipole transmitter for calibration purposes is not shown.}
    \label{fig:station}
\end{figure}

\section{Description of ARIANNA-200 and performance}
Similar in basic design to several stations in the 10 station test-bed array, each ARIANNA-200 station includes 8 antenna channels: 4 log periodic dipole antennas (LPDA) pointing down, 3 LPDAs pointing up, and a dipole (Fig.~\ref{fig:station} left). The LPDAs, which are high-gain broadband receivers, are installed to a depth of only \SI{2}{m}, while the dipole is located \SI{10}{m} below the surface. 
Each station will operate autonomously using solar and wind power and will communicate in almost real time through the Iridium satellite network. 
Based on the ARIANNA test bed experience \cite{ARIANNANuSearch}, there are several options to deploy two hundred ARIANNA stations.  In one method, for example, the deployment team installs all LPDA antennas in rectangular slots created by electrically heated melting devices, which incorporates the key design elements of the cylindrical hole-melter that successfully drilled several holes to the required depth without the need for continuous supervision. Based on the ARIANNA test bed experience, we estimate that two ARIANNA-200 stations can be deployed per day with an 8 person team. The complete ARIANNA-200 array can be deployed in three years. 

The performance and reliability of the ARIANNA architecture was also demonstrated by the ARIANNA test bed array, consisting of 10 ARIANNA stations. It ran successfully from December 2014 to completion of the program in November 2019, achieving operational live-time of 86\% during the sunlit summer months, and a neutrino analysis efficiency of 80\% relative to trigger level at a trigger threshold of 4 times the RMS noise \cite{ARIANNANuSearch}. An experimental prototype of a portable wind generator survived for 2 years and achieved 39\% runtime during periods when sunlight was not available. Incremental changes in the geometry of the wind generator and battery capacity are expected to increase the operational live-time to 70\% during the completely dark winter months \cite{WindTurbineICRC2019}. To summarize: ARIANNA-200 is expected to operate for more than 80\% of the year using non-centralized fuel-free sources of power.  

ARIANNA-200 achieves state-of-the-art sensitivity by optimizing the trigger bandwidth for the high gain LPDA antennas, reaching an equivalent threshold in signal to noise of 2 times the RMS noise. The effective area of ARIANNA-200 (Fig.~\ref{fig:area} left) grows rapidly at neutrino energies above \SI{e17}{eV}, complementing the capabilities of IceCube at lower energies. 

The angular direction of the neutrino is computed from the polarization angle, the arrival direction at the detector and the viewing angle relative the Cherenkov cone. The viewing angle measurement benefits from the large bandwidth of the LPDAs and data acquisition electronics.  ARIANNA-200 will measure the angular direction of nearly every event with an accuracy of \SI{3}{\degree} or better and the energy to within a factor 2 \cite{GlaserICRC2019}. Neutrino energy requires a measurement of the distance to the interaction vertex, which is measured by the powerful DnR technique, to identify the location of the vertex of nearly every event with high precision \cite{DnR2019}. In the DnR method, the distance to the vertex is related to time difference between two signal paths of the radio emission, one that propagates directly to the dipole receiver and the second ray that reflects from the surface to the dipole. Many systematic uncertainties are reduced by observing the time delay in a single dipole. Consequently, ARIANNA test bed studies have shown that the relative precision of the time delay is $<$\SI{0.1}{ns} \cite{DnR2019}. 

Though the baseline reconstruction capabilities have been established by the ARIANNA test bed, we plan additional in-situ calibration campaigns to improve the precision of ice modeling and reduce systematic errors currently limiting the response of the detector.  These goals will be facilitated by continuous monitoring of the snow accumulation.

\section{Practical advantages of architecture and site location}
The Moore’s Bay site is only \SI{110}{km} from McMurdo Station, the largest science base in Antarctica.  The relatively close location provides important logistical flexibility, including the possibility to support construction operations by using overland tracked vehicles to transport cargo.  The sea-level location is generally warmer than higher elevation sites in Antarctica, and the site has better conditions for wind-generated power \cite{WindTurbineICRC2019}.

ARIANNA technologies consume only 5 watts of power per station, which is supplied by solar panels during summer months and wind generators during the continuously dark winter months. The ARIANNA concept avoids the need to deploy (and eventually retrieve) hundreds of kilometers of power and/or communication cables from a central location. The utility of this forward-looking feature will be more evident as the area footprint of the neutrino telescopes increase in future designs. 

Due to advances by the ARIANNA collaboration in event recognition by deep-learning and other proven analysis techniques \cite{ARIANNANuSearch}, high priority neutrino candidates will be transmitted over the reliable Iridium satellite network, which was used routinely in the ARIANNA test bed. Deep learning will be employed to identify neutrino candidates in a matter of seconds, providing a real time alert for the multi-messenger communities. 

The near surface location of the components of the ARIANNA station allows routine maintenance and possibility of technology upgrades, providing the opportunity for ARIANNA to follow the science. The existing infrastructure provides advanced capabilities to implement system upgrades. For example, new trigger software can be uploaded remotely over wireless connections during summer operations. 

\section{Backgrounds}
The protected geography of Moore’s Bay shields ARIANNA-200 from anthropogenic radio-fre\-quency noise created by McMurdo Station, about \SI{100}{km} distant. Backgrounds associated with radio production by cosmic ray collisions in the atmosphere are intrinsically directional and they can be identified by upward facing LPDA antennas \cite{ARIANNACosmicRay2017}. In addition, cosmic ray signals will not produce the characteristic double pulse waveform in the dipole antenna employed by the DnR method for neutrino vertex reconstruction \cite{DnR2019}. Perhaps the most serious background arises from high-energy muons in cosmic ray air showers that penetrate the ice surface and occasionally radiate high energy photons within the ice \cite{Garcia-Fernandez:2020dhb}. The photons initiate an electromagnetic shower in the vicinity of the ARIANNA station that appears identical to a neutrino signal. We thoroughly studied this potential background and found that the expected rate of background events for the full ARIANNA-200 array is less than 0.01 events per year. Thus, for the sensitivity of this pathfinder project, muon background events are negligible even after 10 years of operation. We note that a large part of the background events can be rejected by tagging cosmic-ray air showers and by measuring the muon energy so that this background can be further mitigated if required by a future detector with significantly larger sensitivity.

\section{Summary}
ARIANNA architecture is fully vetted and ready to contribute to the multi-messenger science program by searching for high-energy neutrino emission from more than half the sky. We propose to expand ARIANNA to 200 stations, with the potential to produce transformative science, by measuring the energy and direction of every neutrino candidate, which is vital to unraveling the mystery of cosmic ray acceleration. 

\section{Acknowledgements}
We are grateful to the U.S. National Science Foundation-Office of Polar Programs, the U.S. National Science Foundation-Physics Division (grant NSF-1607719) for granting the ARIANNA test bed array at Moore's Bay. Without the invaluable support of the people at McMurdo, the ARIANNA stations would have never been built.

We acknowledge funding from the German research foundation (DFG) under grants GL 914/1-1 (CG) and NE 2031/2-1 (DGF, ANe, IP, CW) and the Taiwan Ministry of Science and Technology (JN, SHW).  HB acknowledges support from the Swedish Government strategic program Stand Up for Energy.  DB and ANo acknowledge support from the MEPhI Academic Excellence Project (Contract No.  02.a03.21.0005) and the Megagrant 2013 program of Russia, via agreement 14.12.31.0006 from 24.06.2013. AvV acknowledges financial support from the European Research Council (ERC) under the European Union's Horizon 2020 research and innovation programme (Grant No. 646623).

\bibliographystyle{JHEP}
\bibliography{references}

\end{document}